\documentclass[useAMS,usenatbib,referee]{biom}
\usepackage{color}

\def\bSig\mathbf{\Sigma}

\def\Vh{\mbox{vech}}

\def\MSE{\mbox{MSE}}
\def\MISE{\mbox{MISE}}
\def\IN{\small\mbox{IN}}
\def\OOS{\small\mbox{OOS}}
\def\full{\small\mbox{full}}
\def\test{\small\mbox{test}}
\def\joint{\small\mbox{jnt}}
\def\margin{\small\mbox{mg}}
\def\ideal{\small\mbox{ide}}

\def\para{\small\mbox{SAG}}
\def\paraT{\small\mbox{SAG(5)}}

\def\paraF{\small\mbox{SAG(2)}}
\def\true{\small\mbox{true}}
\def\tumor{\small\mbox{tumor}}

\usepackage{mathrsfs,amsmath,amssymb}
\usepackage[ruled]{algorithm2e}
\usepackage{algorithmic}
\usepackage{subfigure} 
\let\TRUE\relax

\usepackage{url}
\usepackage{graphicx}
\usepackage{rotating}
\usepackage{multirow}
\usepackage{comment}
\usepackage{lineno}

\title[Semiparametric Gaussian Mixture Model with Spatial Dependence]{A Semiparametric Gaussian Mixture Model with Spatial Dependence and Its Application to Whole-Slide Image Clustering Analysis}

\author{Baichen Yu$^{1}$, 
Jin Liu$^{2,*}$\email{liujin@nankai.edu.cn. The corresponding author.}, and Hansheng Wang$^{1}$ \\
$^{1}$Guanghua School of Management, Peking University, Beijing, China \\
$^{2}$School of Statistics and Data Science, KLMDASR, LEBPS and LPMC, Nankai University, Tianjin, China}

\begin{document}


\date{}



\pagerange{\pageref{firstpage}--\pageref{lastpage}} 
\volume{}
\pubyear{}
\artmonth{}


\doi{}


\label{firstpage}


\begin{abstract}
We develop here a semiparametric Gaussian mixture model (SGMM) for unsupervised learning with valuable spatial information taken into consideration. Specifically, we assume for each instance a random location. Then, conditional on this random location, we assume for the feature vector a standard Gaussian mixture model (GMM). The proposed SGMM allows the mixing probability to be nonparametrically related to the spatial location. Compared with a classical GMM, SGMM is considerably more flexible and allows the instances from the same class to be spatially clustered. To estimate the SGMM, novel EM algorithms are developed and rigorous asymptotic theories are established. Extensive numerical simulations are conducted to demonstrate our finite sample performance. For a real application, we apply our SGMM method to the CAMELYON16 dataset of whole-slide images (WSIs) for breast cancer detection. The SGMM method demonstrates outstanding clustering performance.\\
\end{abstract}

%



\maketitle


%

\section{Introduction}
\label{s:intro}

This work is primarily motivated by the whole-slide image (WSI) analysis for tumor detection \citep{ghaznavi2013digital}. WSI is a powerful and widely adopted medical imaging technology, which is used for cancer prognosis prediction \citep{lee2022derivation}, cancer classification \citep{breen2025comprehensive}, and metastasis detection \citep{bejnordi2017diagnostic}. 
More specifically, a WSI is one particular type of color image with ultrahigh resolution. For instance, Figure \ref{Fig:DataIllus}(A) demonstrates a WSI example of the sentinel lymph nodes for breast cancer diagnosis. Unfortunately, conventional deep learning based image classification methods cannot be immediately applied to the WSI data due to its extremely large size. One possible solution is to cut the whole image into many sub-images. By treating the WSI sample as a bag and sub-images as instances contained in the bag, various multiple instance learning (MIL) methods can be applied \citep{zhou2018brief}. Nevertheless, the applicability of those MIL methods hinges on one critical condition. That is the WSI-level labels must be provided for both positive and negative (e.g., tumor or non-tumor) WSI samples. They become inapplicable if only one positive WSI sample is given. One possible remedy is to provide sub-image-level labels for the WSI sample. In this case, essentially any appropriate type of classifier can be trained. Unfortunately, this is seldom the case in real practice. Then, how to learn the sub-image-level labels for the sake of tumor localization in a fully unsupervised way becomes a problem of great interest.

To this end, we develop here a fully unsupervised clustering method. Specifically, we first cut a positive WSI sample into many small-sized sub-images (i.e., instances), which are then represented by feature vectors. In this regard, essentially any well-pretrained self-supervised deep learning models can be used \citep{kang2023benchmarking}. Among those models, the UNI model of \cite{chen2024towards} demonstrated the state-of-the-art (SOTA) prediction accuracy on a number of important benchmark datasets. Therefore, we are motivated to apply this SOTA model to our WSI instances for feature extraction in this work. Next, a classical Gaussian Mixture Model (GMM) is imposed on the extracted feature vector for clustering analysis \citep{mclachlan2019finite}. Meanwhile, it has been empirically well documented that the WSI instances are likely to be spatially clustered \citep{ye2019breast}. However, to the best of our knowledge, most existing GMM methods, for example the popularly used spectral estimation, do not take the spatial information into consideration \citep{loffler2021optimality,chen2024achieving}. 
Then, how to modify the classical GMM appropriately so that the valuable spatial information can be effectively incorporated becomes a problem of great interest.

To address this issue, various spatially-informed clustering methods have been developed. These methods can be classified into three categories. The first category contains those Bayesian methods, which treat each instance as a node in a spatial network. Thereafter, it encourages the spatially adjacent instances in the spatial network to share the same label \citep{zhao2021spatial}. Unfortunately, those methods often suffer from expensive computational cost, since various MCMC algorithms have to be used for model estimation. 
The second category contains those methods, which model the mixture component parameters nonparametrically \citep{lee2018nonparametric,zeng2025semiparametric}. As a consequence, those methods allow the mixture component parameters (e.g., means and covariances) from the same class but different locations with large distances to be very different. In contrast, for most WSI analysis, we should expect the mixture component parameters from the same class to be similar regardless of their locations. The third category contains those methods, which impose a GMM structure not only on the feature vector but also on the spatial locations of the instances \citep{zhou2020early}. The consequence is that the marginal distribution of the instance locations becomes another Gaussian mixture. Unfortunately, the empirical distribution of the instance locations on a WSI seems obviously not a Gaussian mixture. 

We are therefore motivated to develop a new GMM method, which allows the mixing probability to be spatially varying in a nonparametric way. Meanwhile, the mixture component parameters (i.e., the means and covariances) are assumed to be constant regardless of the instance locations. For convenience, we refer to the new method as a semiparametric Gaussian mixture model (SGMM). Our main contributions are given as follows. Compared with GMM, SGMM allows valuable spatial information to be taken into consideration. Compared with various Bayesian methods, SGMM is computationally much more efficient. Compared with the fully nonparametric methods, SGMM retains the interpretability of the feature vectors and also the parametric convergence rate of the component parameters. Compared with a fully parametric approach, SGMM allows the mixing probability to be fully nonparametric, which provides more flexibility in modeling capability. To estimate an SGMM, efficient estimation methods and computational algorithms are developed. Rigorous asymptotic theories are established. Extensive numerical simulations and a real WSI data analysis are conducted.

The rest of the article is organized as follows. Section 2 introduces the model setup and the estimation methods. The asymptotic analysis of the theoretical properties is investigated in Section 3. Extensive numerical experiments are presented in Section 4. The article is concluded with a brief discussion in Section 5. All computational details and technical proofs are provided in the Web Appendices.

\section{Model}
\label{s:model}

\subsection{Model and Notations}

Let $({\bf X}_{i},Y_{i},{\bf S}_{i})$ be the data collected from the $i$-th ($1\leq i\leq N$) sub-image of a given WSI sample. Here, $Y_{i}\in\{1,2,\cdots,K\}$ is the class label for the $i$-th sub-image. Note that $Y_{i}$ is not directly observed. Assume the mixing probability $P(Y_{i}=k)=\pi_{k}>0$ for every $1\leq k\leq K$ with $\sum\limits_{k=1}^{K} \pi_{k}=1$. Given $Y_{i}=k$, the feature vector ${\bf X}_{i}=(X_{ij})\in\mathbb{R}^{p}$ and the random location ${\bf S}_{i}=(S_{i1},S_{i2})^{\top}\in\mathbb{S}$ are generated independently \citep{zhou2020early}, where $\mathbb{S}\subset\mathbb{R}^{2}$ is a $2$-dimensional compact domain. Conditional on $Y_{i}=k$, we assume for ${\bf X}_{i}$ a multivariate Gaussian distribution with mean ${\bm\mu}_{k}=(\mu_{kj})\in\mathbb{R}^{p}$ and covariance ${\bm\Sigma}_{k}=(\sigma_{kj_{1}j_{2}})\in\mathbb{R}^{p\times p}$. We further assume that the random location ${\bf S}_{i}={\bf s}$ given $Y_{i}=k$ follows an unknown but continuous distribution with a probability density function given by $g_{k}({\bf s})$ with $g_{k}({\bf s})\geq 0$ for any $1\leq k\leq K$ and ${\bf s}\in\mathbb{S}$.

It is remarkable that we assume the feature vectors and the spatial locations are conditionally independent given the class label $Y_{i}$. However, tumor tissues often exhibit heterogeneity to some extent in their expression across different organ locations. Then, it is more appropriate to assume that ${\bf X}_{i}$ and ${\bf S}_{i}$ are conditionally dependent. Nevertheless, a careful analysis reveals that most of the heterogeneity is due to the fact that different tissues from different locations actually belong to different and further refined sub-categories (instead of only two major classes: tumor vs. non-tumor). For the tissues belonging to the same and sufficiently refined sub-category, we find that their expressions are fairly similar regardless of their locations. Therefore, this conditional independence assumption, similar to the one used by \cite{zhou2020early}, should hold approximately at least for our intended WSI setting, as long as the number of clusters $K$ is not too small.

Write $\phi_{{\bm\mu},{\bm\Sigma}}({\bf x}) = (2\pi)^{-p/2} |{\bm\Sigma}|^{-1/2} \exp\big\{-({\bf x}-{\bm\mu})^{\top}{\bm\Sigma}^{-1}({\bf x}-{\bm\mu})\big/2\big\}$ as the probability density function of a multivariate Gaussian distribution with mean ${\bm\mu}\in\mathbb{R}^{p}$ and covariance ${\bm\Sigma}\in\mathbb{R}^{p\times p}$. Then, the joint probability density function of $({\bf X}_{i},{\bf S}_{i})$ is given by $f({\bf x},{\bf s}) = \sum\limits_{k=1}^{K} \pi_{k}\phi_{{\bm\mu}_{k},{\bm\Sigma}_{k}}({\bf x}) g_{k}({\bf s})$. Write $G({\bf s}) = \sum\limits_{k=1}^{K} \pi_{k}g_{k}({\bf s})$. We then have the conditional probability density function of ${\bf X}_{i}$ with ${\bf S}_{i}={\bf s}\in\mathbb{S}$ given by $f_{{\bf s}}({\bf x}) = f({\bf x},{\bf s}) / G({\bf s}) = \sum\limits_{k=1}^{K} \pi_{k}({\bf s}) \phi_{{\bm\mu}_{k},{\bm\Sigma}_{k}}({\bf x})$, where $\pi_{k}({\bf s}) = \pi_{k} g_{k}({\bf s}) / G({\bf s}) = P(Y_{i}=k|{\bf S}_{i}={\bf s})$. It is interesting to note that the conditional distribution of ${\bf X}_{i}$ with ${\bf S}_{i}$ given (but not $Y_{i}$) remains a Gaussian mixture model. However, the mixing probability changes from $\pi_{k}$ to $\pi_{k}({\bf s})$, which is a nonparametric function in ${\bf s}$. Therefore, we refer to $\pi_{k}({\bf s})$ as the local mixing probability. 

\subsection{The Estimation Methods}

We next consider how to estimate this SGMM model. For an arbitrary symmetric matrix ${\bf B}=(b_{ij})\in\mathbb{R}^{p\times p}$, define a half-vectorization operator as $\Vh({\bf B}) = (b_{ij}:1\leq i\leq j\leq p)\in\mathbb{R}^{p(p+1)/2}$. 
Write ${\bm\theta}_{k} = \big({\bm\mu}_{k}^{\top},\Vh({\bm\Sigma}_{k})^{\top}\big)^{\top}\in\mathbb{R}^{p(p+3)/2}$ for $k=1,\cdots, K$, and ${\bm\Theta} = \big({\bm\theta}_{1}^{\top}, \cdots, {\bm\theta}_{K}^{\top}\big)^{\top}\in\mathbb{R}^{Kp(p+3)/2}$. Note that the marginal density of ${\bf X}_{i}$ is given by $f({\bf X}_{i}) = \sum\limits_{k=1}^{K} f({\bf X}_{i}|Y_{i}=k) P(Y_{i}=k) = \sum\limits_{k=1}^{K} \pi_{k}\phi_{{\bm\mu}_{k},{\bm\Sigma}_{k}}({\bf X}_{i})$. Then, without utilizing the spatial information ${\bf S}_{i}$, the marginal log-likelihood of ${\bf X}_{i}$ can be written as
\begin{equation}\label{X-loglik}
    \mathcal{L}_{\bf x}({\bm \Theta}, {\bm\pi}) = \sum\limits_{i=1}^{N} \log\bigg\{\sum\limits_{k=1}^{K} \pi_{k}\phi_{{\bm\mu}_{k},{\bm\Sigma}_{k}}\Big({\bf X}_{i}\Big)\bigg\}.
\end{equation}
Note that $\sum\limits_{k=1}^{K} \pi_{k} = 1$. Therefore, we can represent the mixing probability by a $(k-1)$-dimensional vector as ${\bm\pi} = \big(\pi_{1}, \cdots, \pi_{K-1}\big)^{\top}\in\mathbb{R}^{K-1}$. Then, the classical maximum likelihood estimator (MLE) can be defined as $\big(\widehat{\bm\Theta}^{\margin}, \widehat{\bm\pi}^{\margin}\big) = \arg\max\limits_{{\bm\Theta}, {\bm\pi}} \mathcal{L}_{\bf x}({\bm\Theta},{\bm\pi})$, for which a standard EM algorithm can be used to compute $\widehat{\bm\mu}_{k}^{\margin}$ and $\widehat{\bm\Sigma}_{k}^{\margin}$ \citep{wu1983convergence}. Here the subscript ``mg" is used to emphasize that $\widehat{\bm\Theta}^{\margin}$ is the MLE computed based on the ``marginal" distribution of ${\bf X}_{i}$ without taking the location information ${\bf S}_{i}$ into consideration. Then a natural question arises: can we further improve the estimation accuracy of $\widehat{\bm\Theta}^{\margin}$ by utilizing the location information ${\bf S}_{i}$ appropriately?

To address this issue, we develop a novel estimator as follows. Let ${\bf s}\in\mathbb{S}$ denote an arbitrary but fixed location. Recall that the conditional density of ${\bf X}_{i}$ given ${\bf s}$ is $f_{s}({\bf X}_{i}) = \sum\limits_{k=1}^{K} \pi_{k}({\bf s}) \phi_{{\bm\mu}_{k},{\bm\Sigma}_{k}}({\bf X}_{i})$. We  then construct a locally weighted likelihood function as 
\begin{equation}\label{S-loglik}
    \mathcal{L}_{\bf s}\big({\bm\Theta}, {\bm\pi}({\bf s})\big) = \sum\limits_{i=1}^{N} \log\bigg\{\sum\limits_{k=1}^{K} \pi_{k}({\bf s})\phi_{{\bm\mu}_{k},{\bm\Sigma}_{k}}\Big({\bf X}_{i}\Big)\bigg\}\mathbb{K}\bigg(\frac{{\bf S}_{i}-{\bf s}}{h}\bigg),
\end{equation}
where $h>0$ is the bandwidth and ${\bm\pi}({\bf s}) = \big(\pi_{1}({\bf s}), \cdots, \pi_{K-1}({\bf s})\big)^{\top} \in\mathbb{R}^{K-1}$ represents the mixing probability locally around $s$ since $\sum\limits_{k=1}^{K} \pi_{k}({\bf s}) = 1$. Moreover, $\mathbb{K}(\cdot)$ is a kernel function defined on $\mathbb{R}^{2}$. Typically, we assume that $\mathbb{K}({\bf s}) = K(s_{1})K(s_{2})$, where $K(t)$ with $t\in\mathbb{R}$ is a probability density function symmetric about $0$. The locally weighted objective function $\mathcal{L}_{\bf s}\big({\bm\Theta}, {\bm\pi}({\bf s})\big)$ involves two types of unknown parameters. They are, respectively, a fixed dimensional parameter $\bm\Theta$ and a nonparametric parameter ${\bm\pi}({\bf s})\in(0,1)^{K-1}$ with infinite dimension. However and fortunately, an initial estimator for $\bm\Theta$ has already been obtained as $\widehat{\bm\Theta}^{\margin}$. We can thus replace $\bm\Theta$ in $\mathcal{L}_{\bf s}\big({\bm\Theta}, {\bm\pi}({\bf s})\big)$ with $\widehat{\bm\Theta}^{\margin}$, which leads to a simplified local loss function as $\mathcal{L}_{\bf s}\big(\widehat{\bm\Theta}^{\margin}, {\bm\pi}({\bf s})\big)$. Here, we adopt the local constant method since the interested parameter $\pi_{k}({\bf s})$ should be bounded in $(0,1)$. Thereafter, an estimator for ${\bm\pi}({\bf s})$ can be obtained as 
$\widehat{{\bm\pi}}({\bf s}) = \arg\max\limits_{{\bm\tau}=(\tau_{k})\in(0,1)^{K-1}}\mathcal{L}_{{\bf s}}(\widehat{\bm\Theta}^{\margin}, {\bm\tau})$. 

To solve optimization \eqref{S-loglik}, we develop a kernel-based EM algorithm (i.e., Algorithm 1) with details given in Web Appendix B. By this algorithm, we should compute $\widehat{\bm\pi}({\bf s})$ for every ${\bf s}\in\big\{{\bf S}_{i}:1\leq i\leq N\big\}$. Consequently, optimization \eqref{S-loglik} needs to be solved for a total of $N$ times. This leads to heavy computation costs. This computational challenge can be well solved by parallel computation, since the computations of $\widehat{\bm\pi}({\bf S}_{i})$s are completely independent of each other. This leads to a significant reduction in computation costs, which is to be shown in the subsequent Section \ref{sec:numerical}. Once $\widehat{{\bm\pi}}({\bf s})$ is computed for every ${\bf s}\in\big\{{\bf S}_{i}:1\leq i\leq N\big\}$, we can then consider how to further upgrade $\widehat{\bm\Theta}^{\margin}$ for better statistical efficiency. 

Recall that the joint probability density of $({\bf X}_{i},{\bf S}_{i})$ is $f({\bf X}_{i},{\bf S}_{i}) = G({\bf S}_{i})\sum\limits_{k=1}^{K} \pi_{k}({\bf S}_{i})  \phi_{{\bm\mu}_{k},{\bm\Sigma}_{k}}({\bf X}_{i})$. Then, a global log-likelihood function can be constructed as 
\begin{align}
    \mathcal{L}^{*}({\bm\Theta},{\bm\Pi}) &= \sum\limits_{i=1}^{N} \log\bigg\{\sum\limits_{k=1}^{K} \pi_{k}\phi_{{\bm\mu}_{k},{\bm\Sigma}_{k}}\Big({\bf X}_{i}\Big) g_{k}\Big({\bf S}_{i}\Big)\bigg\} \notag \\ &= \sum\limits_{i=1}^{N} \log\Bigg[\bigg\{\sum\limits_{k=1}^{K} \pi_{k}\Big({\bf S}_{i}\Big)\phi_{{\bm\mu}_{k},{\bm\Sigma}_{k}}\Big({\bf X}_{i}\Big) \bigg\}G\Big({\bf S}_{i}\Big)\Bigg]\notag \\ 
    & = \sum\limits_{i=1}^{N} \log\bigg\{\sum\limits_{k=1}^{K} \pi_{k}({\bf S}_{i})\phi_{{\bm\mu}_{k},{\bm\Sigma}_{k}}\Big({\bf X}_{i}\Big)\bigg\} + \sum\limits_{i=1}^{N} \log\bigg\{G\Big({\bf S}_{i})\Big)\bigg\}.\label{global-loglik-full}
\end{align}
Next, replace ${\bm\Pi} = \big({\bm\pi}({\bf S}_{1})^{\top},\cdots, \pi({\bf S}_{N})^{\top}\big)^{\top}\in(0,1)^{N(K-1)}$ by $\widehat{\bm\Pi} = \big(\widehat{\bm\pi}^{\top}({\bf S}_{1}),\cdots, \widehat{\bm\pi}^{\top}({\bf S}_{N})\big)^{\top} \in(0,1)^{N(K-1)}$, where $\widehat{\bm\pi}({\bf S}_{i})$ is the pilot estimator obtained by maximizing \eqref{S-loglik}. This leads to a loss function $\mathcal{L}^{*}({\bm \Theta}, \widehat{\bm\Pi})$ in $\bm\Theta$. We can then optimize $\mathcal{L}^{*}({\bm \Theta}, \widehat{\bm\Pi})$ with respect to $\bm\Theta$. This leads to a joint estimator $\widehat{\bm\Theta}^{\joint} = \arg\max\limits_{\bm\Theta} \mathcal{L}^{*}({\bm \Theta}, \widehat{\bm\Pi})$. The subscript ``jnt" is used to emphasize the fact that $\widehat{\bm\Theta}^{\joint}$ utilizes the information from the ``joint" distribution of $({\bf X}_{i},{\bf S}_{i})$. Note that the second term in \eqref{global-loglik-full} depends on $\pi_{k}$ and $g_{k}({\bf S}_{i})$ through $G({\bf S}_{i})$. It has nothing to do with $\bm\Theta$. Therefore, we have $\widehat{\bm\Theta}^{\joint} = \arg\max\limits_{\bm\Theta} \mathcal{L}({\bm \Theta}, \widehat{\bm\Pi})$, where 
\begin{gather}
    \mathcal{L}({\bm\Theta},{\bm\Pi}) = \sum\limits_{i=1}^{N} \log\bigg\{\sum\limits_{k=1}^{K} \pi_{k}({\bf S}_{i})\phi_{{\bm\mu}_{k},{\bm\Sigma}_{k}}\Big({\bf X}_{i}\Big)\bigg\},\label{global-loglik}
\end{gather}
for which only the first term in \eqref{global-loglik-full} is involved. To optimize $\mathcal{L}({\bm \Theta}, \widehat{\bm \Pi})$ with respect to $\bm \Theta$, a standard EM algorithm can be used with a slight modification. Its numerical convergence theory has been well established in the past literature \citep{wu1983convergence,xu1996convergence,balakrishnan2017statistical}. Note that our method is a multi-step estimator that combines parametric and nonparametric estimation methods, which is different from a classical semiparametric efficiency theory.
Further note that our estimation process is executed only once and not iteratively. One can also iterate our optimization problems \eqref{S-loglik} and \eqref{global-loglik} to estimate $\bm\Pi$ and ${\bm\Theta}^{\joint}$ iteratively sufficiently, so that a fully-iterated estimator $\widehat{\bm\Theta}^{\full}$ can be obtained, as long as the time cost is not a serious concern. However, both our subsequent theoretical analysis and numerical simulations in Section \ref{sec:num-simu} show that the efficiency gain in this regard is very limited and is ignorable asymptotically.

\section{Theoretical Properties}\label{sec:thm}

\subsection{Theoretical Properties of $\widehat{\bm\pi}_{k}({\bf s})$}\label{sec:thm-nonpara}

We next study the asymptotic properties of the various estimators. Define ${\bf a}_{k}({\bf s}) = \pi_{k}^{-1}({\bf s}){\bf e}_{k}\in\mathbb{R}^{K-1}$ for $k=1,\cdots,K-1$ and ${\bf a}_{K} = -\pi_{K}^{-1}({\bf s})1_{K-1}\in\mathbb{R}^{K-1}$, where ${\bf e}_{k}\in\mathbb{R}^{K-1}$ denotes the $k$-th column of the $(K-1)$-dimensional identity matrix ${\bf I}_{K-1}\in\mathbb{R}^{(K-1)\times (K-1)}$. In addition, ${\bf 1}_{K-1}\in\mathbb{R}^{K-1}$ refers to the all-one vector with $K-1$ dimensions. For notation convenience, define $\alpha_{k}({\bf X}_{i},{\bf s}) = \alpha_{k}({\bf X}_{i},{\bf s}|{\bm\Theta}) = P(Y_{i}=k|{\bf X}_{i},{\bf S}_{i}={\bf s})$. Let ${\bm\delta} = ({\bm\pi}^{\top}, {\bm\Theta}^{\top})^{\top}\in\Delta\subset\mathbb{R}^{q}$ be the true parameter, where $q = Kp(p+3)/2+K-1$, and $\Delta$ is assumed to be a compact parameter space. We first derive the asymptotic results of the marginal MLE $\widehat{\bm\Theta}^{\margin}$ as follows.
\begin{theorem}\label{thm-1}
    Assume the SGMM model as described in Section 2.1, we have $\sqrt{N}\Big(\widehat{\bm\delta}^{\margin}-{\bm\delta}\Big)\rightarrow_{d} N\big\{0, (\widetilde{\bm\Omega}^{\margin})^{-1}\big\}$ as $N\rightarrow\infty$, where $\widehat{\bm\delta}^{\margin} =  (\widehat{\bm\pi}^{\margin\top},\widehat{\bm\Theta}^{\margin\top})^{\top}\in\mathbb{R}^{q}$, and $\widetilde{\bm\Omega}^{\margin}\in\mathbb{R}^{q\times q}$ is the asymptotic precision matrix defined in Web Appendix F.
\end{theorem}
\noindent The proof of Theorem \ref{thm-1} is provided in Web Appendix F. By Theorem \ref{thm-1}, we know that $\big\|\widehat{\bm\Theta}^{\margin} - {\bm\Theta}\big\| = O_p(1/\sqrt{N})$ as $N\rightarrow\infty$, where $\|{\bf x}\| = \sqrt{{\bf x}^{\top}{\bf x}}$ stands for the usual $\ell_{2}$-norm for an arbitrary vector $\bf x$. We next study the asymptotic properties of $\widehat{\bm\pi}({\bf s})$. To this end, a set of commonly used regularity conditions are necessarily needed. They are given in Web Appendix A with detailed discussions. Then, we have the following theorem.
\begin{theorem}\label{thm-2}
    Under Conditions (C1)-(C4), we know that for any fixed ${\bf s}\in\mathbb{S}\subset\mathbb{R}^{2}$, there exists a local MLE $\widehat{\bm\pi}({\bf s})$ such that: (1) $\big|\widehat{\bm\pi}({\bf s}) - {\bm\pi}({\bf s})\big| = O_{p}(1/\sqrt{Nh^{2}})$; and (2) 
    \begin{gather}
    \sqrt{Nh^{2}}\Big\{\widehat{\bm\pi}({\bf s})-{\bm\pi}({\bf s})\Big\}\rightarrow_{d} N\Big\{0,\eta_{2}^{2}{\bf V}^{-1}({\bf s})\Big\}, \label{thm-nonpara-eq}
    \end{gather}
    where ${\bf V}({\bf s}) = \int \big\{\sum\limits_{k=1}^{K} \alpha_{k}({\bf x},{\bf s}){\bf a}_{k}({\bf s})\big\}\big\{\sum\limits_{k=1}^{K}\alpha_{k}({\bf x}, {\bf s}) {\bf a}_{k}({\bf s})\big\}^{\top} f_{{\bf s}}({\bf x})\mathrm{d}{\bf x}\in\mathbb{R}^{(K-1)\times (K-1)}$.
\end{theorem}
\noindent The detailed proof of Theorem \ref{thm-2} is given in Web Appendix G. By the proof in Web Appendix G, we know that the asymptotic distribution of $\widehat{\bm\pi}({\bf s})$ remains the same, if the pilot estimator $\widehat{\bm\Theta}^{\margin}$ obtained in \eqref{X-loglik} is replaced by the true value $\bm\Theta$. This result is not surprising, since the marginal estimator $\widehat{\bm\Theta}^{\margin}$ is a $\sqrt{N}$-consistent estimator. This is a convergence rate faster than the nonparametric $\sqrt{Nh^{2}}$ convergence rate of $\widehat{\bm\pi}({\bf s})$. Therefore, the estimation error (i.e., the data re-using effect) induced by $\widehat{\bm\Theta}$ is asymptotically ignorable for $\widehat{\bm\pi}({\bf s})$. Moreover, we are able to extend the pointwise convergence result \eqref{thm-nonpara-eq} to a uniform one in the sense that $\sup\limits_{{\bf s}\in\mathbb{S}} \|\widehat{{\bm\pi}}({\bf s})-{\bm\pi}({\bf s})\big\| = O_{p}\big(\sqrt{\log N/(Nh^{2})}\big)$. The technical details are given in Lemma 1 in Web Appendix H.1.

\subsection{Theoretical Properties of $\widehat{\bm\Theta}^{\joint}$}\label{sec:thm-para}

We next study the asymptotic behavior of $\widehat{\bm\Theta}^{\joint}$. We then have the following Theorem \ref{thm-3}, whose detailed proof is provided in Web Appendix H.
\begin{theorem}\label{thm-3}
    Under Conditions (C1)-(C4), we have $\sqrt{N}\Big(\widehat{\bm\Theta}^{\joint}-{\bm\Theta}\Big)\rightarrow_{d} N\Big\{0,\big({\bm\Omega}_{1}^{\joint}\big)^{-1}\\ {\bm\Omega}^{\joint}\big({\bm\Omega}_{1}^{\joint}\big)^{-1}\Big\}$ as $N\rightarrow\infty$, where both ${\bm\Omega}^{\joint}$ and ${\bm\Omega}_{1}^{\joint}\in\mathbb{R}^{q'\times q'}$ are defined in Web Appendix H and $q' = Kp(p+3)/2$.
\end{theorem}
\noindent By Theorem \ref{thm-3}, we know that $\widehat{\bm\Theta}^{\joint}$ remains $\sqrt{N}$-consistent, even if a nonparametric estimator $\widehat{\bm\Pi}$ is involved for computing $\widehat{\bm\Theta}^{\joint}$. To gain some quick insight, we study below some special but also important cases. Specifically, we introduce an ideal estimator $\widehat{\bm\Theta}^{\ideal}$, which is defined in the same way as $\widehat{\bm\Theta}^{\joint}$ with 
$\widehat{\bm\Pi}$ replaced by $\bm\Pi$. We then have the following corollary.
\begin{corollary}\label{corollary-1}
Under Conditions (C1)-(C4) and assume that the local mixing probability $\Pi$ is given, we then have $\sqrt{N}\Big(\widehat{\bm\Theta}^{\ideal}-{\bm\Theta}\Big)\rightarrow_{d} N\Big\{0, \big({\bm\Omega}_{1}^{\joint}\big)^{-1}\Big\}$ as $N\rightarrow\infty$.
\end{corollary}
\noindent The proof of Corollary \ref{corollary-1} is provided in Web Appendix I.1. Moreover, in many real applications, observations from the same class are heavily clustered spatially. Accordingly, the information provided by ${\bf S}_{i}$ should be extremely helpful. To theoretically reflect this interesting phenomenon, we can assume that $E\big[\prod\limits_{k=1}^{K}\{1-\alpha_{k}({\bf X}_{i},{\bf S}_{i})\}\big] \rightarrow_{p} 0$ as $N\rightarrow \infty$. We then have the following corollary. 
\begin{corollary}\label{corollary-2}
    Assume Conditions (C1)-(C4) hold and the local mixing probability $\Pi$ is given. Further assume that $E\big[\prod\limits_{k=1}^{K}\{1-\alpha_{k}({\bf X}_{i},{\bf S}_{i})\}\big] \rightarrow_{p} 0$ as $N\rightarrow \infty$, we then have $\sqrt{N}\Big(\widehat{\bm\Theta}^{\ideal}-{\bm\Theta}\Big)\rightarrow_{d} N\Big\{0, \big({\bm\Omega}_{2}^{\joint}\big)^{-1}\Big\}$ as $N\rightarrow\infty$, where ${\bm\Omega}_{2}^{\joint}\in\mathbb{R}^{q'\times q'}$ are defined in Web Appendix I.2. It can be verified that ${\bm\Omega}_{2}^{\joint} - {\bm\Omega}_{1}^{\joint}$ is positive semi-definite.
\end{corollary}
\noindent The proof of Corollary \ref{corollary-2} is also provided in Web Appendix I.2. By Corollary \ref{corollary-2}, we know that, under the situation that the data are heavily clustered spatially, the ideal estimator $\widehat{\bm\Theta}^{\ideal}$ should be statistically more efficient. In this case, the latent class label $Y_{i}$s can be recovered by the posterior probability perfectly in the sense that $\alpha_{k}({\bf X}_{i},{\bf S}_{i}) - I(Y_{i}=k) \rightarrow_{p} 0$. Therefore, the ideal estimator $\widehat{\bm\Theta}^{\ideal}$ reduces to the simple moment estimator with the best efficiency. We can then reasonably conclude that the joint estimator $\widehat{\bm\Theta}^{\joint}$ should be more efficient than the marginal estimator $\widehat{\bm\Theta}^{\margin}$, as long as the spatial information is strong enough.

\section{Numerical Experiments}\label{sec:numerical}

\subsection{Simulation Studies}\label{sec:num-simu}

To demonstrate the finite sample performance of the different estimating methods, we present here a number of simulation studies. 

{\sc Study 1.} The objective of this study is to compare the statistical efficiency of the initial MLE $\widehat{\bm\Theta}^{\margin}$ and the joint estimator $\widehat{\bm\Theta}^{\joint}$. For $\widehat{\bm\Theta}^{\margin}$, we apply the $K$-means algorithm to obtain an initial estimator. For $\widehat{\bm\Theta}^{\joint}$, both the $K$-means estimator and $\widehat{\bm\Theta}^{\margin}$ are considered as the initial estimators. We fix the number of classes $K=2$, the feature means ${\bm\mu}_{1} = -{\bm\mu}_{2} = {\bf 1}_{p}\in\mathbb{R}^{p}$ with $p\in\{2,5,10\}$, and ${\bm\Xi}_{p} = (\rho_{ij})$ with $\rho_{ij} = 0.5^{|i-j|}$. 
Set ${\bm\Sigma}_{1} = 16 {\bm\Xi}_{p}$, ${\bm\Sigma}_{2} = 9 {\bm\Xi}_{p}$, and $\pi_{1} = 0.4$. 
The underlying distribution of the spatial location conditional on the class label is also set to be a Gaussian distribution with mean ${\bm\mu}_{S1} = -{\bm\mu}_{S2} = (1,1)^{\top}$ and covariance matrix ${\bm\Sigma}_{S} = 0.5{\bm\Xi}_{2}$ but truncated on $[-5,5]^{2}$. 
The bandwidth $h$ is specified to be $h = 2.5\times N^{-1/3}$. For a given $N$, we randomly replicate the experiment for a total of $R=500$ times and obtain $R$ mean squared error (MSE) values for every target parameter (i.e., ${\bm\mu}_{1},{\bm\mu}_{2},{\bm\Sigma}_{1}$ and ${\bm\Sigma}_{2}$). The MSE values of ${\bm\mu}_{1}$ are log-transformed, averaged, and then reported in Table \ref{Tab:Simu-res1}. The averaged computational time and the numbers of iterations are also reported. The MSE values of other parameters are qualitatively similar and are presented in Web Appendix D.1.
\begin{sidewaystable}[!ht]
\centering
\caption{Log-transformed mean squared errors for estimating ${\bm\mu}_{1}$ using different estimators. The estimators in the quote refer to the initialization methods. The column $\log(\MSE)$ reports the log-transformed MSE values. The column Time refers to the average CPU time in seconds. The column Iteration is the average numbers of iterations. OOS refers to out-of-sample prediction of the local mixing probability.}
\label{Tab:Simu-res1}
\resizebox{\textwidth}{!}{\begin{tabular}{ccrrrrrrrrrrrr}
\hline
\multicolumn{2}{c}{$p$}                                                                                                                              & \multicolumn{4}{c}{$2$}                                                                                                   & \multicolumn{4}{c}{$5$}                                                                                                   & \multicolumn{4}{c}{$10$}                                                                                                  \\
\multicolumn{2}{c}{$n$}                                                                                                                              & \multicolumn{1}{c}{$500$} & \multicolumn{1}{c}{$1{,}000$} & \multicolumn{1}{c}{$2{,}000$} & \multicolumn{1}{c}{$5{,}000$} & \multicolumn{1}{c}{$500$} & \multicolumn{1}{c}{$1{,}000$} & \multicolumn{1}{c}{$2{,}000$} & \multicolumn{1}{c}{$5{,}000$} & \multicolumn{1}{c}{$500$} & \multicolumn{1}{c}{$1{,}000$} & \multicolumn{1}{c}{$2{,}000$} & \multicolumn{1}{c}{$5{,}000$} \\ \hline
\multirow{3}{*}{$\log(\MSE)$}                                                           & $\widehat{\bm\Theta}^{\margin}$                               & $-0.996$                  & $-1.644$                      & $-2.296$                      & $-3.389$                      & $-1.259$                  & $-1.981$                      & $-2.877$                      & $-4.006$                      & $-1.585$                  & $-2.601$                      & $-3.581$                      & $-4.587$                      \\
                                                                                        & $\widehat{\bm\Theta}^{\joint}$ ($K$-means)                    & $-1.193$                  & $-1.880$                      & $-2.860$                      & $-3.804$                      & $-1.623$                  & $-2.521$                      & $-3.501$                      & $-4.528$                      & $-2.066$                  & $-3.100$                      & $-4.062$                      & $-5.014$                      \\
                                                                                        & $\widehat{\bm\Theta}^{\joint}$ ($\widehat{\bm\Theta}^{\margin}$) & $-1.184$                  & $-1.880$                      & $-2.860$                      & $-3.804$                      & $-1.626$                  & $-2.522$                      & $-3.501$                      & $-4.527$                      & $-2.081$                  & $-3.100$                      & $-4.062$                      & $-5.014$                      \\ \hline
\multirow{3}{*}{Time (s)}                                                               & $\widehat{\bm\Theta}^{\margin}$                               & $0.49$                    & $0.83$                        & $1.44$                        & $3.19$                        & $0.25$                    & $0.46$                        & $0.71$                        & $1.41$                        & $0.20$                    & $0.27$                        & $0.36$                        & $0.93$                        \\
                                                                                        & $\widehat{\bm\Theta}^{\joint}$ ($K$-means)                    & $0.06$                    & $0.07$                        & $0.09$                        & $0.23$                        & $0.08$                    & $0.08$                        & $0.11$                        & $0.23$                        & $0.09$                    & $0.09$                        & $0.12$                        & $0.29$                        \\
                                                                                        & $\widehat{\bm\Theta}^{\joint}$ ($\widehat{\bm\Theta}^{\margin}$) & $0.01$                    & $0.02$                        & $0.02$                        & $0.02$                        & $0.03$                    & $0.02$                        & $0.02$                        & $0.03$                        & $0.04$                    & $0.02$                        & $0.02$                        & $0.03$                        \\ \hline
\multirow{3}{*}{Iter}                                                                   & $\widehat{\bm\Theta}^{\margin}$                               & $706.5$                   & $964.7$                       & $1193.5$                      & $1231.8$                      & $271.9$                   & $378.2$                       & $375.0$                       & $325.0$                       & $164.7$                   & $170.4$                       & $129.6$                       & $117.7$                       \\
                                                                                        & $\widehat{\bm\Theta}^{\joint}$ ($K$-means)                    & $26.1$                    & $22.3$                        & $17.6$                        & $14.4$                        & $36.1$                    & $20.2$                        & $13.9$                        & $11.1$                        & $45.2$                    & $18.3$                        & $11.9$                        & $9.9$                         \\
                                                                                        & $\widehat{\bm\Theta}^{\joint}$ ($\widehat{\bm\Theta}^{\margin}$) & $19.4$                    & $17.0$                        & $12.9$                        & $9.6$                         & $32.7$                    & $17.0$                        & $10.1$                        & $7.4$                         & $43.6$                    & $15.6$                        & $8.7$                         & $6.8$                         \\ \hline
\multirow{3}{*}{$\widehat{\pi}_{k}({\bf s})$}                                                 & $\log(\MISE)$                                              & $-2.278$                  & $-2.660$                      & $-3.262$                      & $-3.899$                      & $-2.575$                  & $-3.374$                      & $-4.208$                      & $-4.904$                      & $-2.928$                  & $-4.104$                      & $-5.069$                      & $-5.668$                      \\
                                                                                        & Time (s)                                                   & $1.23$                    & $2.36$                        & $5.25$                        & $19.92$                       & $1.15$                    & $2.47$                        & $5.35$                        & $19.26$                       & $1.04$                    & $2.44$                        & $5.18$                        & $20.28$                       \\
                                                                                        & Iter                                                       & $65.2$                    & $34.9$                        & $19.9$                        & $82.2$                        & $56.1$                    & $35.0$                        & $111.9$                       & $76.7$                        & $51.0$                    & $145.7$                       & $98.0$                        & $61.6$                        \\ \hline
\multirow{3}{*}{%
        $\begin{array}{c} 
            \widehat{\pi}_{k}(\mathbf{s}) \\ 
            \text{(OOS)} 
        \end{array}$%
    } & $\log(\MISE)$                                              & $-2.250$                  & $-2.674$                      & $-3.261$                      & $-3.917$                      & $-2.555$                  & $-3.398$                      & $-4.219$                      & $-4.941$                      & $-2.902$                  & $-4.126$                      & $-5.112$                      & $-5.723$                      \\
                                                                                        & Time (s)                                                   & $1.28$                    & $2.41$                        & $5.22$                        & $19.90$                       & $1.22$                    & $2.50$                        & $5.28$                        & $19.56$                       & $1.09$                    & $2.43$                        & $5.15$                        & $20.63$                       \\
                                                                                        & Iter                                                       & $64.8$                    & $34.6$                        & $19.7$                        & $82.2$                        & $55.3$                    & $34.5$                        & $112.5$                       & $76.9$                        & $51.4$                    & $146.0$                       & $98.4$                        & $61.8$                        \\ \hline
\end{tabular}}
\end{sidewaystable}

By Table \ref{Tab:Simu-res1}, we find that, with a given feature dimension $p$, both the marginal estimator $\widehat{\bm\Theta}^{\margin}$ and the joint estimator $\widehat{\bm\Theta}^{\joint}$ are statistically consistent, as their log-transformed MSE values steadily decrease as the sample size increases. For example, with $p=10$, the $\log(\MSE)$ of $\widehat{\bm\Theta}^{\joint}$ with $K$-means initialization decreases from $-2.066$ to $-5.014$ as $N$ grows from $500$ to $5{,}000$. Furthermore, we find that the MSE values of the joint estimator are much lower than those of the marginal estimator, which implies that the joint estimator is statistically more efficient. For example, with $p=10$ and $N=5{,}000$, the $\log(\MSE)$ of $\widehat{\bm\Theta}^{\joint}$ with $K$-means initialization is $-5.014$, which is much smaller than that of $-4.587$ of $\widehat{\bm\Theta}^{\margin}$. This confirms our theoretical results in Section 3. The results associated with different feature dimensions are qualitatively similar. Moreover, we find that different initial estimators for $\widehat{\bm\Theta}^{\joint}$ (i.e., $K$-means and $\widehat{\bm\Theta}^{\margin}$) lead to very similar estimation results. This suggests that our iterative algorithm is fairly insensitive to the choice of the initial parameter. We next compute the average CPU time consumed by different estimators. By Table \ref{Tab:Simu-res1}, we find that the computational time and the average number of iterations can be significantly reduced by using $\widehat{\bm\Theta}^{\margin}$ as an initial estimator. This is expected, since $\widehat{\bm\Theta}^{\margin}$ is an initial estimator more accurate than $K$-means. Additionally, the mean integrated square errors (MISEs) of the nonparametric estimator $\widehat{\pi}_{k}({\bf s})$ on both the in-sample and out-of-sample data points are also reported in Table \ref{Tab:Simu-res1}. Specifically, the in-sample MISE is defined to be $\MISE_{\IN} = R^{-1}\sum\limits_{r=1}^{R} N^{-1}\sum\limits_{i=1}^{N}(\widehat{\pi}_{1}({\bf S}_{i}) - \pi_{1}({\bf S}_{i}))^{2}$, and the out-of-sample (OOS) MISE is defined to be $\MISE_{\OOS} = R^{-1}\sum\limits_{r=1}^{R} N^{-1}\sum\limits_{i=1}^{N}(\widehat{\pi}_{1}({\bf S}_{i}^{*}) - {\pi}_{1}({\bf S}^{*}_{i}))^{2}$, where ${\bf S}_{i}^{*}$s are an independent copy of ${\bf S}_{i}$s. By Table \ref{Tab:Simu-res1}, we find that $\log(\MISE)$ for both in-sample and out-of-sample data points steadily decreases as sample size $N$ increases. This confirms our theoretical claims in Theorem 2.

{\sc Study 2.}  The objective of this study is to compare the finite sample performance of our method with that of the spatially-aided GMM (SAG) of \cite{zhou2020early} in terms of estimation accuracy. The SAG model assumes another independent Gaussian mixture distribution for the spatial location ${\bf S}_{i}$, conditional on the class label $Y_{i}$. 
In this study, we follow the data generating process of \cite{zhou2020early} and set the total number of components to be $L_{\true} = 5$. For the actual parameter estimation, different working numbers of mixture components (denoted by $L$) are used (e.g., $L=2$). Denote the resulting estimators of the SAG model with different working numbers of components as $\widehat{\bm\Theta}^{\para(L)}$. In each replicate, 4 different estimators are compared. They are, respectively, $\widehat{\bm\Theta}^{\margin}$, $\widehat{\bm\Theta}^{\paraF}$, $\widehat{\bm\Theta}^{\paraT}$, and $\widehat{\bm\Theta}^{\joint}$. Then, we randomly replicate the experiment for a total of $R=500$ times. 
\begin{figure}[!ht]
    \centering
    \includegraphics[width=0.85\textwidth]{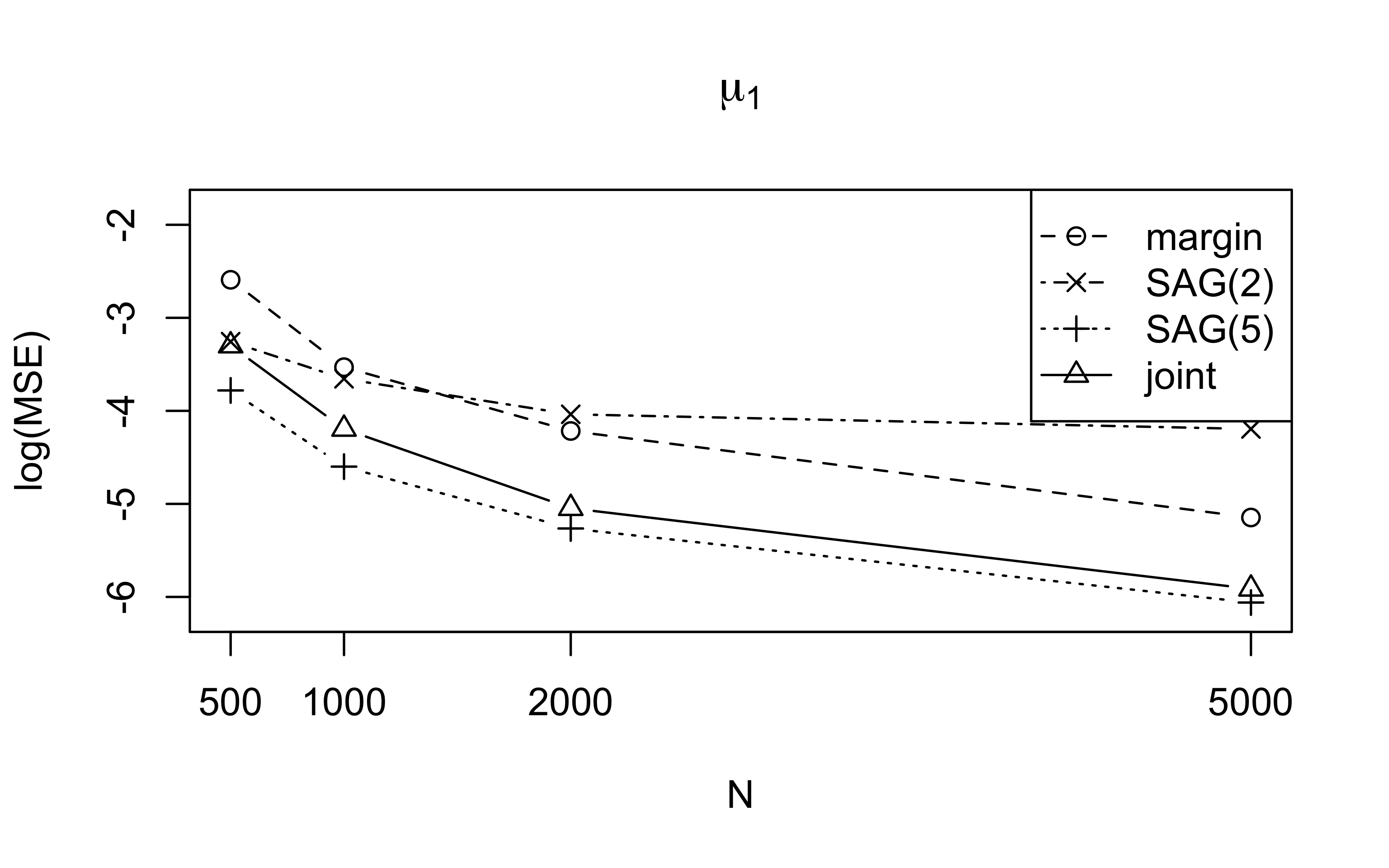}
    \caption{Log-transformed mean squared errors for ${\bm\mu}_{1}$ using different estimating methods under the SAG model \citep{zhou2020early} when $L_{\true}=5$.} 
    \label{Fig:Simu-res2}
\end{figure}
Next, the MSE values for estimating ${\bm\mu}_{1}$ are log-transformed, averaged, and plotted in Figure \ref{Fig:Simu-res2}. By Figure \ref{Fig:Simu-res2}, we find that the MSE values of both $\widehat{\bm\Theta}^{\joint}$ and $\widehat{\bm\Theta}^{\paraT}$ are uniformly lower than those of $\widehat{\bm\Theta}^{\margin}$. Therefore, both the estimators $\widehat{\bm\Theta}^{\joint}$ and the correctly specified SAG estimator $\widehat{\bm\Theta}^{\paraT}$ can improve the efficiency of the estimation by incorporating the spatial information. Moreover, we find that our joint estimator $\widehat{\bm\Theta}^{\joint}$ is comparable to the correctly specified fully parametric estimator $\widehat{\bm\Theta}^{\paraT}$ as the sample size is large enough. 
However, when the number of components is mis-specified as $L=2$, the SAG estimator becomes inconsistent $\widehat{\bm\Theta}^{\paraF}$. The MSE values of other parameters are qualitatively similar and are presented in Web Appendix D.1.

{\sc Study 3.} The objective of this study is to demonstrate the clustering performance of the proposed methods in terms of clustering accuracy. We follow the same setting as the previous {\sc Study 2} and compare the clustering performance of the $4$ different estimators. They are, respectively, the classical GMM, SAG(2), SAG(5), and the SGMM. For each clustering method and each random replication, various evaluation metrics (i.e., AUC, IoU, and ARI) are computed; see Web Appendix C.2 for the technical details. Their average values are then recorded in Table \ref{Tab:Study3}. Note that this experiment is based on the setting of $K=2$. The additional explorations of multiple values of $K>2$ are provided in Web Appendix D.2.
\begin{table}[!h]
\centering
\caption{The averaged prediction results using different estimating methods with different evaluation metrics.}
\label{Tab:Study3}
\begin{tabular}{cccccc}
\hline
Metric               & N    & GMM  & SAG(2) & SAG(5) & SGMM \\ \hline
\multirow{4}{*}{AUC} & 500  & 76.6 & 92.7   & 97.9   & 97.8 \\
                     & 1{,}000 & 78.0 & 92.9   & 98.7   & 99.1 \\
                     & 2{,}000 & 78.6 & 93.4   & 99.1   & 99.6 \\
                     & 5{,}000 & 78.8 & 93.1   & 99.3   & 99.7 \\ \hline
\multirow{4}{*}{IoU} & 500  & 49.7 & 74.7   & 84.2   & 85.0 \\
                     & 1{,}000 & 50.3 & 73.7   & 87.1   & 89.6 \\
                     & 2{,}000 & 50.5 & 75.2   & 88.7   & 92.2 \\
                     & 5{,}000 & 50.4 & 74.4   & 89.7   & 93.7 \\ \hline
\multirow{4}{*}{ARI} & 500  & 24.7 & 57.2   & 74.3   & 75.8 \\
                     & 1{,}000 & 25.3 & 55.5   & 79.2   & 83.1 \\
                     & 2{,}000 & 25.5 & 57.7   & 81.8   & 87.4 \\
                     & 5{,}000 & 25.4 & 55.6   & 83.4   & 89.8 \\ \hline
\end{tabular}
\end{table}
By Table \ref{Tab:Study3}, we find that all the four methods improve as the sample size $N$ increases. The methods of SAG(2), SAG(5), and SGMM all outperform the classical GMM clearly, since the valuable spatial information is not taken into consideration by the GMM. Moreover, recall that the true model is generated from SAG(5). We find that the result of our SGMM method is fairly comparable to that of the SAG(5). The actual performance of the SGMM could be even better if the sample size $N$ is large enough. However, when the model is mis-specified as SAG(2), our SGMM outperforms it significantly in terms of AUC, IoU, and ARI. This further illustrates our model flexibility in terms of clustering analysis.

\begin{figure}[!ht]
    \centering
    \includegraphics[width=0.85\textwidth]{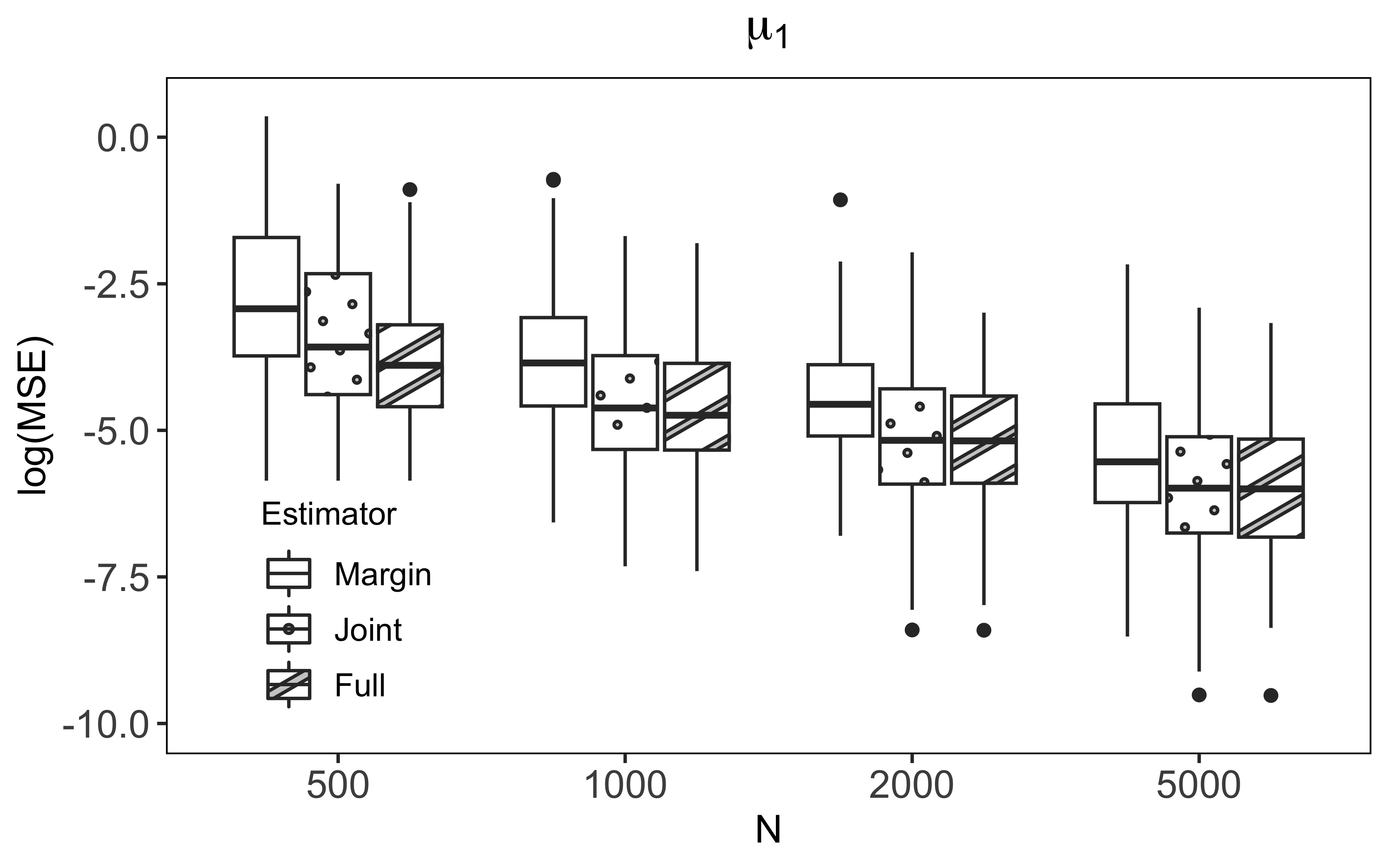}
    \caption{Log-transformed mean squared errors for ${\bm\mu}_{1}$ obtained by iterating once and fully iterating.} 
    \label{Fig:Simu-res4}
\end{figure}

{\sc Study 4.} Note that given the initial estimator, $\widehat{\bm\Theta}^{\joint}$ is a one-iteration estimator in terms of subsequently optimizing \eqref{S-loglik} and \eqref{global-loglik}. Next, we can then take $\widehat{\bm\Theta}^{\joint}$ as the initial estimator to repeat the optimization process \eqref{S-loglik} and \eqref{global-loglik} iteratively. We can repeat this process for a sufficient number of times till the algorithm numerically converges. This leads to a fully-iterated estimator $\widehat{\bm\Theta}^{\full}$. Then, the $\log(\MSE)$s of various estimators (i.e., $\widehat{\bm\Theta}^{\margin}$, $\widehat{\bm\Theta}^{\joint}$, and $\widehat{\bm\Theta}^{\full}$) are calculated and compared by boxplots. The simulation setup is the same as that of {\sc Study 1}. The detailed results for estimating ${\bm\mu}_{1}$ are then reported in Figure \ref{Fig:Simu-res4}. The MSE values of other parameters are qualitatively similar and are presented in Web Appendix D.1. 
By Figure \ref{Fig:Simu-res4}, we find that the finite sample performance of $\widehat{\bm\Theta}^{\joint}$ can be further improved by $\widehat{\bm\Theta}^{\full}$. However, the relative improvement margin decreases steadily and disappears eventually as the sample size increases. This is expected, since both $\widehat{\bm\Theta}^{\joint}$ and $\widehat{\bm\Theta}^{\full}$ share the same asymptotic efficiency due to the following reasons. Recall that $\widehat{\bf\Theta}^{\margin}$ is a global parameter estimator and hence is $\sqrt{N}$-consistent, while $\widehat{\bm\pi}({\bf s})$ is a local estimator and is $\sqrt{Nh^{2}}$-consistent. Consequently, the estimation error due to $\widehat{\bf\Theta}^{\margin} - {\bf\Theta}$ (i.e., the data reusing effect) is asymptotically negligible for $\widehat{\bm\pi}({\bf s}) - \bm\pi({\bf s})$. Therefore, the asymptotic distribution of $\widehat{\bm\pi}({\bf s})$ remains the same as the estimator computed with the true parameter ${\bf\Theta}$ given. Then, the asymptotic efficiency of $\widehat{\bm\pi}({\bf s})$ cannot be further improved by iteratively improving the estimation accuracy of ${\bf\Theta}$. Since the asymptotic efficiency of $\widehat{\bm\pi}({\bf s})$ cannot be further improved by additional iteration, that of $\widehat{\bf\Theta}^{\joint}$ cannot be further improved asymptotically either.

\subsection{Real Data Analysis}

\begin{figure}[!ht]
    \centering
    \includegraphics[width=0.95\textwidth]{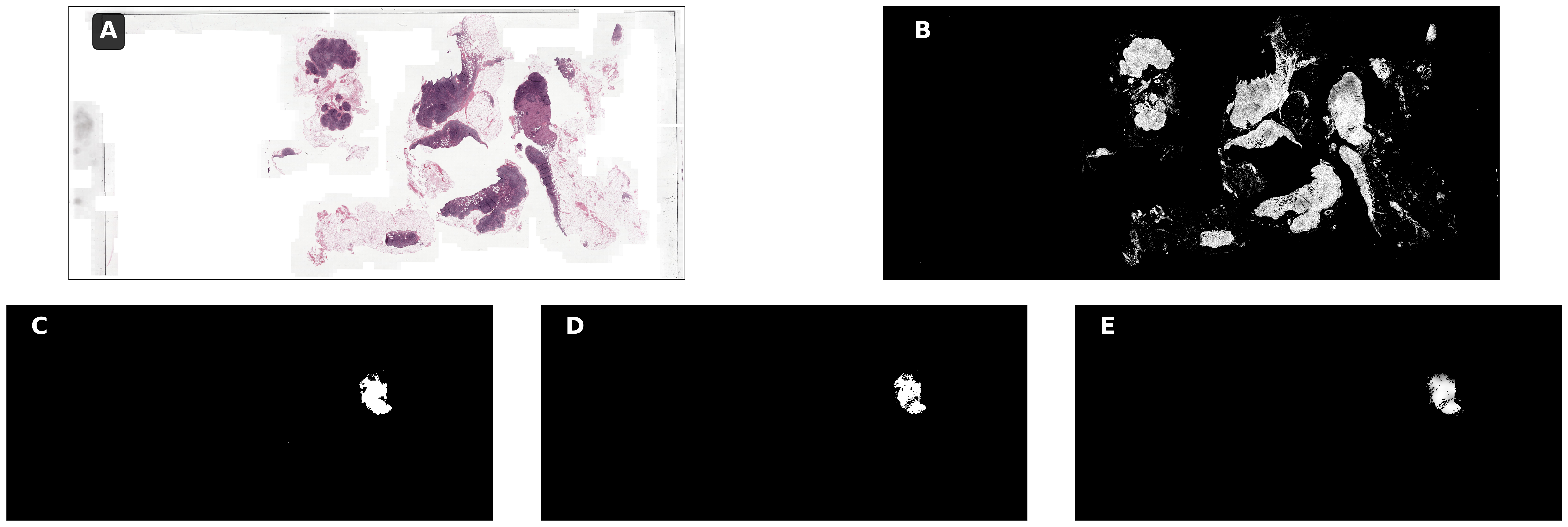}
    \caption{The prediction result of an arbitrarily selected WSI sample. A, the original WSI. B, the tissue region of the WSI. The bright area displays the informative region obtained by the method of \protect\cite{otsu1979threshold}. C, the human-annotated ground truth of the tumor region. D, the posterior probability of being a tumor by our SGMM. E, the estimated local mixing probability by our SGMM.}
    \label{Fig:DataIllus}
\end{figure}

To demonstrate the practical usefulness of the proposed methods, we present here a real data example. The dataset used here is the CAMELYON16 dataset, which is an important benchmark dataset about breast carcinoma metastasis detection  \citep{bejnordi2017diagnostic}. We follow \cite{wang2022label} and select WSIs with medium-size tumors (i.e., lesion ratio greater than $0.1$ and smaller than $0.9$). This leads to a total of $11$ WSIs. For a quick understanding, an arbitrarily selected WSI sample is illustrated in Figure \ref{Fig:DataIllus}(A).

This is an image with an ultrahigh resolution of $97{,}792 \times 221{,}184 = 2.2\times 10^{10}$ pixels. Therefore, appropriate data preprocessing is necessarily needed. In this regard, we follow a widely accepted standard procedure in the literature \citep{lee2022derivation,breen2025comprehensive}. This preprocessing process contains a total of 3 important steps. They are, respectively, (1) cutting a WSI into a number of sub-images with size $256\times 256$; (2) using the method of \cite{otsu1979threshold} to discard the background tiles and obtain Figure \ref{Fig:DataIllus}(B); and (3) extracting features for every sub-image by a deep learning method. Here, we apply the UNI model of \cite{chen2024towards}, which is already pre-trained on a massive $77$TB dataset by self-supervised contrastive learning. This leads to a $d = 512$ dimension feature vector. 
We then apply the principle component analysis on this feature vector so that its dimension can be further reduced to dimension $p$. Different values of dimension $p\in\{2,5,10\}$ are studied subsequently.

We next apply the proposed SGMM method to those instances. To this end, the number of clusters needs to be specified. Recall that the goal here is to distinguish between tumor and non-tumor regions. Therefore, it is natural to set the number of clusters to $K=2$. However, as we mentioned before in Section \ref{s:model}, the tumor cells and normal cells can be further classified into more refined sub-categories, since different refined sub-categories vary in morphology and hence statistical distribution. If we directly set $K=2$, the tumor related sub-categories can possibly be confused with those non-tumor related ones. This often leads to sub-optimal classification accuracy. Therefore, we are motivated to consider a relatively larger range of choices for the number of clusters. This leads to more refined sub-category discovery. More specifically, we have tried various $K\in\{2,\cdots,8\}$. Accordingly, the joint estimator $\widehat{\bm\Theta}^{\joint}$ can be computed and the posterior probabilities $\widehat{\alpha}_{k}({\bf X}_{i},{\bf S}_{i})$ can be evaluated. For a robust evaluation, we compared their average performance on different dimension $p$s and different WSI samples as follows.

Note that for this particular dataset, the human-annotated ground truth $Y_{i}^{\tumor}\in\{0,1\}$ is only provided for two major classes (i.e., $Y_{i}^{\tumor}=1$ for tumor and $Y_{i}^{\tumor}=0$ otherwise) as shown in Figure \ref{Fig:DataIllus}(C). To compare our clustering results with this human-annotated label, we have to convert the refined $K$ sub-categories into two major classes (i.e., tumor v.s. normal). The technical details for this converting process are given in Web Appendix C.1. By this converting process, we obtain for each instance an integrated probability of being a tumor. Matching the predicted probability with the human-annotated ground truth $Y_{i}^{\tumor}$, we are able to compute the AUC \citep{ling2003auc} and Intersection-over-Union \citep[IoU,][]{rezatofighi2019generalized}. We also compute the Adjusted Rand Index \citep[ARI,][]{hubert1985comparing}. The technical details of those evaluation metrics are also introduced in Web Appendix C.2.

Our SGMM method is then compared with a total of 3 competing methods in terms of out-of-sample prediction accuracy. The first method is the classical GMM, which makes no use of spatial information. The other two competitors are the SAG method of \cite{zhou2020early} and the BayesSpace method of \cite{zhao2021spatial}. Both methods are spatially-informed. The detailed results are given in Table \ref{Tab:Real-aver}.
\begin{table}[!h]
\centering
\caption{The averaged prediction results for different $K$ values by using four competing methods. The top performance is highlighted in boldface for each row. The average time cost is reported in parentheses in seconds.}
\label{Tab:Real-aver}
\begin{tabular}{cccccc}
\hline
Metric               & K & GMM         & SAG                  & BayesSpace    & SGMM                 \\ \hline
\multirow{7}{*}{AUC} & 2 & 93.0 (2.1)  & 93.6 (25.6)          & 82.7 (947.1)  & \textbf{94.5} (2.5)  \\
                     & 3 & 89.8 (3.2)  & \textbf{93.8} (41.3) & 86.7 (939.8)  & 91.1 (3.7)           \\
                     & 4 & 88.8 (5.8)  & \textbf{92.6} (66.1) & 90.6 (950.8)  & 92.2 (6.5)           \\
                     & 5 & 90.2 (8.2)  & \textbf{93.2} (82.4) & 92.9 (947.4)  & 93.0 (9.2)          \\
                     & 6 & 90.8 (10.1) & 93.2 (109.0)         & 93.2 (946.8)  & \textbf{93.7} (11.3) \\
                     & 7 & 91.4 (16.3) & 92.3 (120.6)         & 93.6 (948.3)  & \textbf{94.1} (27.7) \\
                     & 8 & 91.0 (30.3) & 92.8 (158.6)         & 93.7 (948.3)  & \textbf{93.9} (32.1) \\ \hline
\multirow{7}{*}{IOU} & 2 & 53.2        & 51.3                 & 50.5          & \textbf{53.7}        \\
                     & 3 & 55.7        & \textbf{60.3}        & 58.0          & 56.2                 \\
                     & 4 & 60.3        & 61.2                 & 59.9          & \textbf{61.3}        \\
                     & 5 & 61.0        & \textbf{63.2}        & \textbf{63.2} & 62.4                 \\
                     & 6 & 62.5        & 63.2                 & 62.6          & \textbf{63.8}        \\
                     & 7 & 62.3        & 63.6                 & \textbf{64.3} & 63.6                 \\
                     & 8 & 62.0        & 63.6                 & \textbf{64.3} & 63.1                 \\ \hline
\multirow{7}{*}{ARI} & 2 & 60.2        & 56.3                 & 48.5          & \textbf{63.4}        \\
                     & 3 & 41.0          & 42.0                   & 34.9          & \textbf{42.7}        \\
                     & 4 & 29.8        & 30.5                 & 28.5          & \textbf{31.1}        \\
                     & 5 & 24.4        & \textbf{27.2}        & 24.2          & 27.0                   \\
                     & 6 & 20.8        & 22.6                 & 20.2          & \textbf{22.9}        \\
                     & 7 & 18.1        & 19.1                 & 17.1          & \textbf{20.1}        \\
                     & 8 & 16.1        & 17.4                 & 14.9          & \textbf{18.5}        \\ \hline
\end{tabular}
\end{table}
By Table \ref{Tab:Real-aver}, we find that GMM performs worst to a large extent for all three performance metrics. This is expected, since no spatial information is used by GMM. Among the other three methods, our SGMM method performs best in most cases. Specifically, we obtain the top performance on $4$ cases in AUC, $3$ cases in IoU, and $6$ cases in ARI. Moreover, we find that usually better IoU results can be obtained by a relatively larger number of clusters $K$, since there may exist some sub-categories in the WSI data. To summarize, we find that our SGMM method seems to be very competitive in terms of various prediction measures as compared with its competitors. More importantly, it offers a key advantage in computational efficiency. Specifically, the averaged computational time of SAG and BayesSpace is about $86$ seconds and $947$ seconds, respectively. On the other side, our SGMM only takes $12$ seconds on average. In contrast, human annotation by a well-trained expert takes about 15 minutes for one single WSI on average \citep{bejnordi2017diagnostic}. The computational advantage of our method is mainly because the kernel-based EM algorithm can be executed in a fully parallel way by tensor-based GPU computing systems; see Web Appendix B for implementation details. For a graphical illustration, we take the WSI of Figure \ref{Fig:DataIllus} as an example. We provide in Figure \ref{Fig:DataIllus}(D) and Figure \ref{Fig:DataIllus}(E) not only the estimated mixing probability but also the estimated posterior probability by SGMM. We find that our SGMM method can recover the tumor region more smoothly and continuously. Practically, pathologists can overlay the heatmaps produced by SGMM on the original WSI as an interactive layer. Consequently, the pathologists can view histological images and our prediction results simultaneously. Therefore, the pathologists can be more focused on the suspicious area for more accurate and refined tumor diagnosis.

\section{Concluding Remarks}

To conclude this article, we discuss here some interesting topics for future study. The SGMM is a semiparametric model, which contains a parametric component (i.e., a Gaussian component). Then, how to relax this parametric assumption for better flexibility is a problem of great interest. Second, our method includes nonparametric smoothing estimation. Therefore, it suffers from both the boundary issue and the curse of dimensionality. Many useful boundary bias reduction techniques, including pseudo data augmentation \citep{chen2012testing} and boundary-modified kernels \citep{su2017time} might be helpful to some extent. Although WSIs are on $2$-dimensional planes, theoretically, it remains interesting to study the curse of dimensionality. One possible solution is to slice a high dimensional WSI sample into multiple $2$-dimensional slices, analyze them separately, and integrate them together using appropriate ensemble methods \citep{tong2019improving}. Another possible solution is to leverage the strength of deep neural networks, which have been shown to have excellent capability in counteracting the curse of dimensionality \citep{jiao2023deep}. Third, our current theoretical results only support feature vectors of a fixed dimension. How to allow the features with high dimension is another future topic. In this regard, appropriate model structures, such as the sparsity assumption \citep{cai2019chime} and the factor structure \citep{zhao2018regularized}, might be helpful.


\backmatter




\section*{Supplementary Materials}

Web Appendices A–I, and data and code referenced in Section \ref{sec:numerical} are available with this paper at the Biometrics website on Oxford Academic.

\section*{Funding}

Jin Liu's research is supported by the National Natural Science Foundation of China (No.12201316). Hansheng Wang's research is supported by the National Natural Science Foundation of China (No.12271012 and 72495123).



\section*{Data Availability}

The whole-slide image (WSI) data that support the findings in this paper are available at \url{https://camelyon16.grand-challenge.org}.


%

\bibliographystyle{biom} 
\bibliography{reference}






\label{lastpage}

\end{document}